\documentclass[aip,apl,reprint]{revtex4-1}

\usepackage{graphicx}
\usepackage{dcolumn}
\usepackage{bm}
\usepackage{natbib}

\begin{document}

\title{Reduction of low-frequency 1/f noise in Al-AlO$_x$-Al tunnel junctions by thermal annealing}

\author{J. K. Julin}
\author{P. J. Koppinen}
\author{I. J. Maasilta}
\affiliation{%
Nanoscience Center, Department of Physics, University of Jyv\"askyl\"a, P. O. Box 35, 
FIN--40014 University of Jyv\"askyl\"a, Finland.
}

\date{\today}

\begin{abstract}

We report that annealing Al--AlO{$_x$}--Al tunnel junctions in a vacuum chamber at temperature of $400^{\circ}$C reduces the characteristic 1/f noise in the junctions, in some cases by an order of magnitude. Both ultra high vacuum and high vacuum fabricated samples demonstrated a significant reduction in the 1/f noise level. Temperature dependence of the noise was studied between 4.2 and 340 Kelvin, with a linear dependence below 100 K, but a faster increase above. The results are consistent with a model where the density of charge trapping two level-systems within the tunneling barrier is reduced by the annealing process.
\end{abstract}

\maketitle


Tunnel junctions are versatile components, which have been used widely as radiation detectors\cite{enss}, superconducting quantum interference (SQUID) magnetometers \cite{book}, single electron transistors and pumps \cite{NATO}, normal metal-insulator-superconductor (NIS) tunnel junction coolers and thermometers\cite{SINIS}, magnetic tunnel junction memory \cite{mood} and superconducting qubits \cite{squbit2}, for example. Thus, improvement of their characteristics can have a wide impact in many applications. By far the most common barrier material used is AlO$_x$ due to its reasonably good properties, ease of fabrication (thermal oxidation at room temperature) and compatibility with superconducting Al. However, the standard AlO$_x$ based junctions are not ideal and typically show aging (slow increase of tunneling resistance) due to glassy dynamics of interfacial electronic traps or other type of two-level systems with wide distribution of relaxation times \cite{nesbitt,koppinen,tan}. Previously, we have shown \cite{koppinen} that vacuum thermal annealing can speed up this aging process significantly and produce stable junctions with improved DC characteristics. Nevertheless, for most of the above applications it is not only the DC characteristics that are important, but also the intrinsic noise properties of the junctions, as excess low-frequency 1/f noise could limit the performance of the device. This is especially true for superconducting qubits, as low-frequency 1/f noise of the critical current leads to dephasing of all types of qubits \cite{vanharlingen,martinisnam,ithier,wellstood,astafiev}. The critical current noise spectral density $S_{I_0}$, furthermore, is widely accepted to to be related to the resistance noise spectral density $S_R$ (measured here) by $S_{I_0}/I_0^2=S_R/R^2$.
In addition, if the noise is produced by charged fluctuators, decoherence also results by a direct electic coupling between the fluctuator and the qubit \cite{martinis}. It is thus quite clear that the quality of the tunnel junction is critical for coherent superconducting circuits \cite{oh}.         

In this paper, we have studied how vacuum thermal annealing affects the intrinsic low-frequency 1/f resistance noise of submicron Al--AlO$_x$--Al tunnel junctions. As most models of the ubiquitous 1/f noise involve a distribution of two-level systems such as charge traps or disorderd atomic positions as the microscopic source of noise  \cite{duttahorn,RevModPhys.60.537} (tunnel junctions are discussed in Refs. \cite{rogers,wakai}),  it is reasonable to assume that the annealing process could also lower the 1/f noise in tunnel junctions, if it improves the DC characteristics \cite{koppinen}. This is indeed true; here we have observed in some cases an order of magnitude reduction in the 1/f noise power density (depending on the quality of the as-fabricated junction) after vacuum annealing. The annealed resistance noise spectral density obtained is about an order of magnitude below that of the recent 1/f noise measurements in slightly larger  Al--AlO$_x$--Al junctions \cite{2006ApPhL..89l2516E}.




Dozens of Al--AlO$_x$--Al tunnel junctions of size $\sim 0.1 \mu \mathrm{m}^2$ (Al film thickness 50 - 100 nm) were fabricated on nitridized or oxidized silicon wafers using electron-beam lithography and two-angle e-beam evaporation of Al (rate 1-2 $\mathrm{\AA}/s$), in either high vacuum (HV) $\sim 10^{-6}$ mbar or ultra-high vacuum (UHV) $\sim 10^{-8}$ mbar conditions. The tunnel barriers were formed by room temperature thermal oxidation in pure oxygen atmosphere, in the HV evaporator at 10 mbar pressure for 4 minutes and in the UHV evaporator at 200 mbar for 4 minutes. Before any metal deposition, the chip was cleaned with $O_2$ plasma at 30 W power in a reactive ion etcher with a pressure of 40 mtorr and a flow of 50 cm$^3$/min, to reduce the effect of PMMA resist contamination.  After the deposition, post-oxidation was used to protect the junctions from unwanted adsorption of contaminants.

The fabrication typically resulted in room temperature tunneling resistances of about 10--20 k$\Omega$ for the UHV samples, while for HV fabricated samples the tunneling resistances were 3--4 times greater. As the size of the junctions was kept constant, the only variables causing the differences in the observed tunneling resistances are the barrier properties, which are known to be sensitive functions of the oxidation conditions. The substrate (SiO or SiN) had no observable effect on the tunneling resistance. 


The annealing process used was the same as described in Refs. \cite{koppinen,konferenssi}. Briefly, the samples were inserted into the opening of a tubular boron nitride resistive heating element  located in a high vacuum chamber. The heater was always set to a temperature of 600 \ensuremath{^\circ}C, as measured by a thermocouple inside the tube. The sample stage was connected to a manipulation rod, which could be moved in and out of the heater, allowing for a quick radiative heating of the sample while inside the tube (no physical contact). The temperature of the sample stage was monitored continuously after the insertion with another thermocouple, so that after the wanted maximum sample stage temperature was reached, a pull-out of the sample could be performed. The cooling of the sample took place in the cold part of the vacuum chamber, slowly in $\sim 1$ hour.

 The maximum  annealing temperature the samples survived was found to be around 400 \ensuremath{^\circ}C, which always produced stable, fully aged junctions for both the HV and UHV fabricated junctions, in agreement with our previous results, where only samples fabricated in HV were studied \cite{koppinen}. 
However, the aging behaviour was  found  to be different between the HV and UHV samples, with slower aging seen for the UHV samples, as expected by simple purity arguments.  
The observed tunneling resistance increases after the 400 \ensuremath{^\circ}C annealing process varied between 10--45\% for the UHV fabricated samples, and 200--300\% for the HV fabricated samples.

\begin{figure}[htbt]
	\centering
		\includegraphics[width=8.5cm]{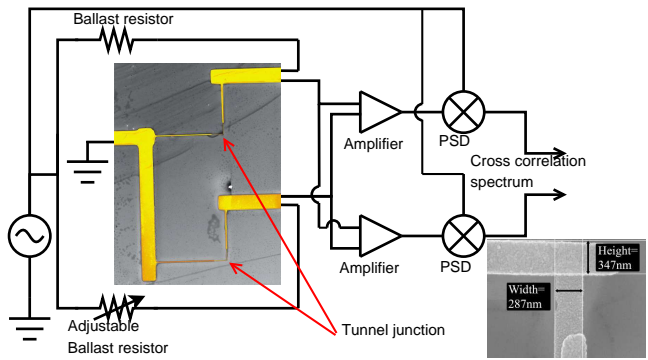}
	\caption{Schematic of the AC modulation bridge noise measurement setup with two pre- and lock-in amplifiers (PSD) and a cross-correlation spectrum analyzer, with a scanning electron (SEM) micrograph of the actual two-junction sample geometry.  The fixed ballast resistor has a resistance of 1 M$\Omega$, and the adjustable resistor (General Radio 1433B) is used to balance the bridge.  Due to the bridge measurement technique both tunnel junctions are measured together. Inset: An SEM micrograph of a typical junction area.}
	\label{fig:acbridge}
\end{figure}

\begin{figure}[htbt]
	\centering	\includegraphics[width=8.5cm]{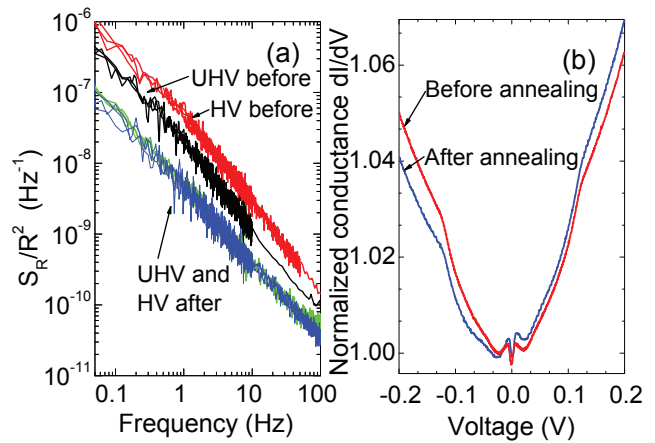}
	\caption{(a) Room temperature resistance noise spectral densities of four double tunnel junction samples fabricated in HV and UHV before ($R_T^{HV}=49 \& 59 \, \mathrm{k} \Omega$, $R_T^{UHV}=23 \& 21 \, \mathrm{k} \Omega$) and after ($R_T^{HV}=155 \& 172 \, \mathrm{k} \Omega$, $R_T^{UHV}=33 \& 34 \, \mathrm{k} \Omega$) annealing at $400^{\circ}$C. The data is normalized with $R_T^2$. After annealing, the spectra are well fitted by $S_R/R^2=0.45\cdot 10^{-8} f^{-1.05}$ 1/Hz. (b) Conductance spectrum of a UHV sample  before ($R_T=12 \, \mathrm{k} \Omega$) and after ($R_T=18 \, \mathrm{k} \Omega$) annealing, demonstrating minor changes in it. The sharp dip around $V=0$ is due to Coulomb blockade.}
	\label{kohinat}
\end{figure}

\begin{figure}[htbt]
	\centering	\includegraphics[width=8.5cm]{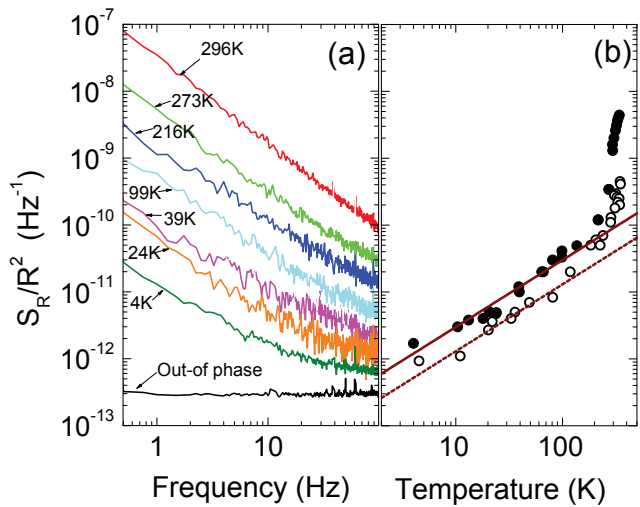}
	\caption{(a) Temperature dependence of the normalized 1/f resistance noise in non-annealed tunnel junctions (data from two samples with $R_T=19 \& 18 \mathrm{k} \Omega$).  The lowest spectrum is the out-of phase component measured at 4 K demonstrating the baseline of the measurement setup. (b) Noise level at 10 Hz s as a function of temperature. Solid circles represent the non-annealed tunnel junctions ($R_T=19 \& 18 \, \mathrm{k} \Omega$), and open circles the annealed junctions ($R_T=4 \& 8 \, \mathrm{k} \Omega$). The lines are fits to linear temperature dependence $S_R/R^2\mathrm{(10 Hz)}=AT$ with $A=1.3\cdot 10^{-13}$ and $A=3 \cdot 10^{-13}$ 1/(HzK). }
	\label{temperatures}
\end{figure}


Measuring of 1/f noise requires a sensitive technique that can resolve the true sample noise below a larger background noise level. We have used the well known AC bridge modulation technique \cite{scofield:985}, which can avoid the high low-frequency voltage preamplifier 1/f noise by shifting the measurement band into the lowest noise frequency region of the preamplifiers, typically around 1 kHz. This is achieved by driving the circuit with a sinusoidal excitation signal at $f \sim 1$ kHz, and using a lock-in amplifier to detect and demodulate the noise back to the original frequency band (see Fig. \ref{fig:acbridge}). By balancing the bridge with the adjustable ballast resistor, the excitation is not measured directly, only noise.  
In addition, we measure the noise using two channels of pre- (Ithaco 1201) and lock-in amplifiers (Stanford Research Systems SR830), and finally record only the cross-correlation spectrum in a two-channel spectrum analyzer (Agilent 89410A) to reduce the background noise level due to cables and preamplifiers even further.  



The effectiveness of the setup was checked by measuring the voltage noise of typical 2 k$\Omega$ resistors, which do not possess significant 1/f noise, and by comparing the results to the theoretically estimated Johnson noise spectral density $S_V=4k_BTR$. At room temperature, the measured noise $v_n \sim 6$ nV/$\sqrt{\mathrm{Hz}}$ was found to match precisely with theory at all measured frequencies 0.1 Hz-100 Hz. Thus, the source of the measured noise was confirmed to emerge only from the sample in this case.  At 4.2 K, the measured white noise level 1.7 nV/$\sqrt{\mathrm{Hz}}$  exceeded the theoretical Johnson noise level by 1.5 nV/$\sqrt{\mathrm{Hz}}$, giving us an estimate for the limits of the contributions from the setup. 

As the 1/f noise in tunnel junctions is generated by resistance fluctuations \cite{rogers,vanharlingen}, its level in voltage units depends on the excitation current. Higher excitation will lead to higher noise level, however, it cannot be increased without limit because of problems with junction breakdown and heating. We found that a 100 nA excitation current was a sufficient compromise so that the junction  voltages were $\sim$ 1 mV and heating powers $\sim 0.1$ nW, causing no problems even at 4.2 K. The measured voltage noise spectral density $S_V$ was converted to resistance noise spectral density $S_R$ (units $\Omega^2/Hz$) by $S_R(f)/R^2=S_V(f)/V^2$,
where  $R$ and $V$ are the sample resistance and voltage, respectively. $S_R$ is expected to scale with resistance as $R^2$, if one assumes the model of resistance fluctuations caused by fluctuations in the effective area (charge traps blocking part of the tunneling area, for example) \cite{rogers,vanharlingen,2006ApPhL..89l2516E}. Thus, to compare our results to this standard model, it is useful to plot all our results scaled as $S_R/R^2$, in which case the noise spectra are expected to be independent of both excitation and sample resistance.


Figure \ref{kohinat} (a) shows typical room temperature noise spectra plotted as scaled resistance noise density $S_R/R^2$, for four different samples, both before and after annealing. By scaling with the sample resistance, the noise spectra results from the same oxidation chamber become identical, but differ between the oxidation chambers (HV or UHV) before annealing. This is consistent with the picture of effective area fluctuations, and with the idea that the areal density of two-level states in the junction depends on the fabrication conditions. Comparing with previous work  \cite{2006ApPhL..89l2516E}, our HV junctions seem to have roughly the same, but UHV junctions slighly smaller noise level \cite{note}. The noise level after annealing at $400^{\circ}$C  was found to be lowered in every sample studied, and, interestingly, attained a common value for both HV and UHV fabricated samples (Fig. \ref{kohinat}). In the HV samples the reduction in scaled resistance noise was as high an order of magnitude. The substrate material (either oxidized or nitridized Si) was found to have no effect on the noise level.

In addition, conductance as a function of voltage also gives information on the characteristics of tunnel junctions. The conductance spectrum can be used to interpret the barrier properties, e.g. resonance peaks are usually caused by unwanted impurities states within the barrier, while the conductance of a perfect barrier should be smooth with a parabolic shape at low voltages. In Ref. \cite{koppinen} we showed that the annealing treatment discussed here removed all excess conductance peaks from the spectrum of HV fabricated samples. In Fig. \ref{kohinat} (b), we show the same comparison for UHV fabricated junctions before and after annealing. We see that there are no excess resonances to begin with, and the annealing treatment only seems to shift the minimum of the parabola, which corresponds to changing the barrier asymmetry. Thus, we have evidence that the higher 1/f noise level seen in our HV samples is correlated with the resonances seen in the conductance spectra, and that these states can be removed by the annealing.

The temperature dependence of the noise spectra was also studied down to 4.2 K, with representative results shown in Fig. \ref{temperatures} (a). In Fig.  \ref{temperatures} (b) we plot the noise at $f=10$ Hz as a function of temperature. In the low temperature range $T < 100$ K, the temperature dependence is roughly linear, and then much faster at higher temperatures in agreement with Ref. \cite{2006ApPhL..89l2516E}, for both as-fabricated and annealed junctions. This low temperature linear temperature dependence is in agreement with the simplest two-level system models \cite{duttahorn}, but in contrast with the $T^2$ dependence found in direct measurements of the critical current or charge noise in superconducting junctions of different material systems \cite{wellstood,vanharlingen,astafiev}.  

Recently, a model was put forward that attempts to explain the difference between normal state and superconducting noise data \cite{faoro}, with the prediction that the superconducting state noise mechanism is distinct from the simpler normal state two-level system model. Because of this possibility, it is not yet clear how our annealing procedure will lower the noise also in the superconducting state. 

In summary, excess 1/f noise can be significantly reduced in Al-AlOx-Al tunnel junctions by vacuum thermal annealing. Many applications of tunnel junctions could possibly benefit from the obtained performance increase.   

This work has been supported by the Academy of Finland under projects 128532 and 118231. We thank J. Pekola for helpful comments.


\begin{thebibliography}{25}
\bibitem [{not()}]{note}%
  \BibitemOpen
  \href@noop {} {\enquote {\bibinfo {title} {If one factors in the expected
  junction area dependence \cite{vanharlingen} $S_R \sim 1/A$, and the fact
  that our junctions have $\sim 1.7$ smaller junction area, even better
  relative noise performance was seen in our junctions.}}\ }\BibitemShut
  {NoStop}%
\end{thebibliography}
\end{document}